\documentclass{article}
\usepackage[utf8]{inputenc}
\usepackage{subcaption}
\usepackage{multirow}
\usepackage{graphicx}
\usepackage{marvosym}
\usepackage{paralist}
\usepackage{hyperref}
\usepackage{amsmath}
\usepackage{authblk}
\usepackage{colortbl}
\usepackage{xcolor}
\usepackage[style=numeric]{biblatex}

\usepackage{float}	
\usepackage{listings}
\usepackage{tabularx}
\usepackage{booktabs}
\usepackage{todonotes}
\usepackage{algorithm}
\usepackage{adjustbox}
\usepackage[noend]{algpseudocode}

\newcommand{\furl}[1]{\footnote{\scriptsize \url{#1}}} 
\usepackage[style=numeric]{biblatex}
\bibliography{references}
\lstdefinestyle{turtle}{%
    morekeywords={a, @prefix},
    morecomment=[s][\textit]{"}{"},
}

\definecolor{Gray}{gray}{0.85}
\definecolor{LightCyan}{rgb}{0.88,1,1}

\newcolumntype{a}{>{\columncolor{Gray}}c}
\newcolumntype{b}{>{\columncolor{white}}c}
\usepackage{academicons}
\definecolor{orcidlogocol}{HTML}{A6CE39}

\begin{document}
\title{Analyzing a Knowledge Graph of Industry~4.0 Standards}
\author[1]{Irl\'an Grangel-Gonz\'alez}
\author[2,3]{Maria-Esther Vidal}
\affil[1]{Robert Bosch GmbH, Corporate Research, Reningen, Germany}
\affil[2]{L3S Research Center, Leibniz University of Hannover, Germany}
\affil[3]{TIB Leibniz Information Centre for Science and Technology, Hannover, Germany}
\affil[ ]{\normalsize\texttt{irlan.grangelgonzalez@de.bosch.com, maria.vidal@tib.eu}}
\date{}
\maketitle

\paragraph{\bf Abstract}
Realizing smart factories according to the Industry~4.0 vision requires intelligent human-to-machine and machine-to-machine \emph{communication}. 
To achieve this goal, components such as actuators, sensors, and cyber-physical systems along with their data, need to be described; moreover, interoperability conflicts arisen from various semantic representations of these components demand also solutions. 
To \emph{empowering communication} in smart factories, a variety of standards and standardization frameworks have been proposed. 
These standards enable the description of the main properties of components, systems, and processes, as well as interactions between them. 
Standardization frameworks classify, align, and integrate industrial standards according to their purposes and features. 
Various standardization frameworks have been proposed all over the world by industrial communities, e.g., RAMI4.0 or IICF. 
While being expressive to categorize existing standards, standardization frameworks may present divergent classifications of the same standard. 
Mismatches between standard classifications generate semantic interoperability conflicts that negatively impact the effectiveness of communication in smart factories.   
In this article, we tackle the problem of standard interoperability across different standardization frameworks, and devise a knowledge-driven approach that allows for the description of standards and standardization frameworks into an Industry 4.0 knowledge graph (I40KG). 
The \emph{STO} ontology represents properties of standards and standardization frameworks, as well as relationships among them. 
The I40KG integrates more than 200 standards and four standardization frameworks. 
To populate the I40KG, the landscape of standards has been analyzed from a semantic perspective and the resulting I40KG represents knowledge expressed in more than 200 industrial related documents including technical reports, research articles, and white papers. 
Additionally, the I40KG has been linked to existing knowledge graphs and an automated reasoning has been implemented to reveal implicit relations between standards as well as mappings across standardization frameworks. 
We analyze both the number of discovered relations between standards and the accuracy of these relations. 
Observed results indicate that both reasoning and linking processes enable for increasing the connectivity in the knowledge graph by up to 80\%, whilst up to 96\% of the relations can be validated. 
These outcomes suggest that integrating standards and standardization frameworks into the I40KG enables the resolution of semantic interoperability conflicts, empowering the \emph{communication} in smart factories. 
\maketitle

\section{Introduction}
\begin{figure*}
\begin{subfigure}{0.49\textwidth}
\includegraphics[width=\linewidth]{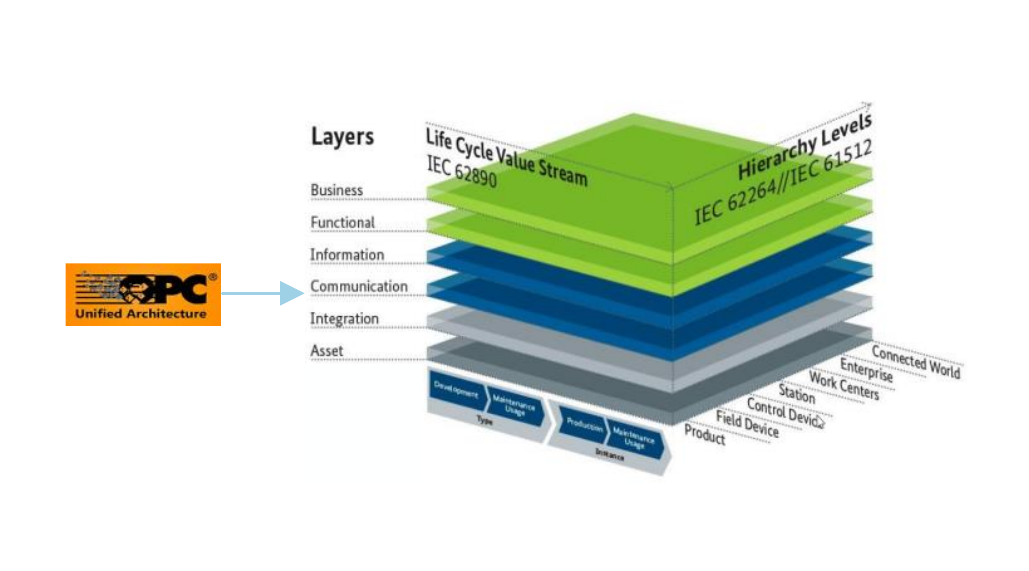}
\caption{RAMI4.0 IT dimension}
\label{fig:RAMIStandards}
\end{subfigure}
\hspace*{\fill}
\begin{subfigure}{0.49\textwidth}
\includegraphics[width=\linewidth]{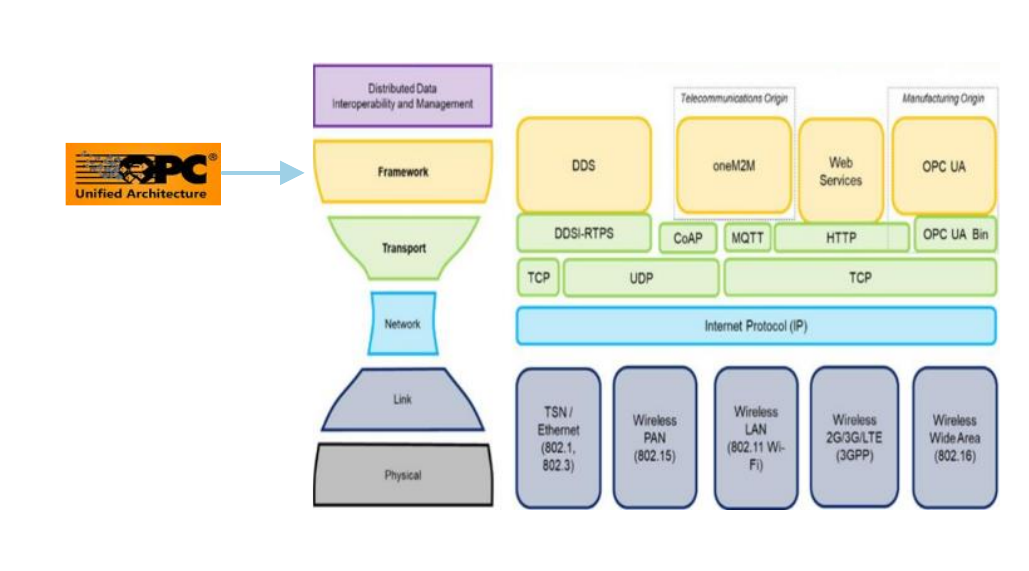}
\caption{IICF Model}
\label{fig:ISA95Pyramid}
\end{subfigure}
\caption{\textbf{Standardization frameworks aligned with I40 Standards}. (a)  RAMI4.0 IT (adjusted from~\cite{ramiModel}). (b) IICF Model (adjusted from~\cite{ramiirareport2017}).
OPC unified architecture (\emph{OPC UA}) standardizes machine-to-machine communication, and it is positioned at different levels in RAMI4.0 IT and IICF. 
OPC UA is at the Framework level in IICF, while in RAMI4.0 IT, it is positioned at communication level and presented as a standard for the description of data management and analytic processes. Thus, the same standard is categorized differently from two standardization frameworks that target the domain of I40.} \label{fig:reference-architectures}
\end{figure*} 
The dynamics of today's world impose new challenges to enterprises.
Globalization, ubiquitous presence of communication networks, new human-machine collaboration scenarios, as well as the development of complex information systems are some of the developments that provoke changes in various areas of industry and society. 
In the engineering and manufacturing domains, there is currently an atmosphere of departure to a new era of digitized production.
This \emph{fourth industrial revolution} has been coined as Industrie~4.0 (I40) in Germany, while related terms, e.g., Industrial Internet, Smart Manufacturing, Industrie du Future, are used to denote similar visions.
They all describe the application of modern IT concepts in industrial contexts, e.g., Internet of Things (IoT), Cyber-Physical Systems (CPS), Internet of Services (IoS), and data-driven architectures.
Following the Industry~4.0 vision, smart factories require to support intelligent human-to-machine and machine-to-machine \emph{communication}~\cite{weyer2015towards}.
To achieve this goal, components such as actuators, sensors, and CPS and their main properties require to be described not only syntactically but also semantically.
More importantly, interoperability conflicts among these components need to be resolved. 
Standards and standardization frameworks have been proposed with the aim of \emph{empowering communication} within smart factories. 
Standards enable the description of the properties of components, systems, and processes, as well as interactions among them. 
Standardization frameworks classify, align, and integrate industrial standards according to their purposes and features. 
Various standardization frameworks have been proposed all over the world by industrial communities, e.g., RAMI4.0 or IICF. 
While being expressive to categorize and align existing standards, standardization frameworks may present divergent interpretations and  classifications of the same standard. 
For instance, OPC UA\furl{https://en.wikipedia.org/wiki/OPC_Unified_Architecture} (OPC Unified Architecture) is classified by RAMI4.0 as a communication standard, while IICF classifies OPC UA as a data acquisition standard.
Although standardization frameworks aim to facilitate interoperability between standards, they still suffer from interoperability conflicts.
Existing approaches have tackled these challenges from various angles, including: 
a) landscaping standards and standardization frameworks with focus on product life-cycles~\cite{LuMF15}; 
b) recognizing similarities and differences between standardization frameworks and describing mappings between areas sharing similar functions ~\cite{ramiirareport2017}. 
However, these methods do not consider the meaning of properties and relations among the standards and standardization frameworks, preventing thus from a seamless integration and interoperability. 
This article tackles this problem and presents a knowledge-driven approach to support semantic interoperability.
We define the problem and approach as follows:
\textbf{Problem.}
We investigate the problem of semantic interoperability between standards and standardization frameworks in the context of Industry~4.0, and identify their main interoperability conflicts and conditions for solving them.
\textbf{Approach.}
We propose a novel knowledge-driven approach that allows for the description of standards and standardization frameworks into the Industry~4.0 Knowledge Graph (I40KG). 
I40KG provides building blocks for discovering new relations of smart factory standards as well as for their semantic mapping to the different standardization frameworks classifying standards world-wide.
The I40KG relies on the \emph{STO} ontology which formally describes the meaning of I40 standards, their properties and relationships.
The semantics represented in \emph{STO} are exploited for the discovery of relations between standards. 
Traversing the I40KG allows for collecting known properties of existing I40 standards, as well as for uncovering relations that are not explicitly stated, e.g., a relation between AML and IEC 61499~\cite{pang2010iec}. 
Moreover, the I40KG can be linked to existing knowledge graphs, such as DBpedia, facilitating the discovery of properties and relations.
We empirically evaluate the quality and accuracy of our proposed approach.
Specifically, we study the performance and accuracy of link discovery across standards and standardization frameworks.
Observed results indicate that the connectivity among standards and standardization frameworks is increased by up to 80\%, while 96\% of the discovered relations are validated.
\textbf{Contributions.}
This article extends our previous work~\cite{Grangel-Gonzalez17}, where an initial version of the I40KG is presented.
Hence, the contributions of this work can be summarized as follows:
\begin{inparaenum}
    \item A methodology to collect and integrate standards and standardization frameworks into a KG. 
    \item An extension of the Standards Ontology (STO)~\cite{Grangel-Gonzalez17} to describe standards and standardization frameworks.
    \item A knowledge graph for Industry~4.0 (I40KG), containing the semantic descriptions for nearly 200 standards and more than 25 standard organizations. 
    \item An empirical evaluation of the quality and accuracy of the integration techniques followed during the creation of I40KG. Observed results provide evidence of soundness of the discovered relations explicitly represented in I40KG, i.e., up to 96\% of the discovered  relations are valid, and connectivity is increased by up to 80\%.
\end{inparaenum}

\section{Preliminaries}
\label{background}
Standardization frameworks classify standards in different layers according to their functions.
In this section, we present the most relevant standardization frameworks and reference architectures to this work.
Figure~\ref{fig:reference-architectures} depicts two reference architectures, RAMI4.0 and IICF, as well as the classification of existing I40 standards into layers. 
It is important to note that some standards, e.g., OPC UA, may be classified at different layers within these architectures.

\subsection{The Reference Architecture Model for Industrie~4.0}
RAMI4.0 is a three-dimensional model that describes fundamental aspects of Industrie~4.0~\cite{ramiModel}.
Figure~\ref{fig:RAMIStandards} illustrates the RAMI4.0 layers and the connections between IT, manufacturers/plants and the product life cycle in a three-dimensional space.
Each dimension shows a particular part of these domains divided into different layers.
The vertical axis to the left represents the IT perspective, comprising layers ranging from the physical device (\emph{asset}) to complex functions as they are available in ERP systems (\emph{functional}) and mappings to \emph{business} models.
The horizontal axis on the left indicates the product life-cycle where \emph{Type} and \emph{Instance} are distinguished as the two main concepts.
The horizontal axis on the right organizes the locations of the features and responsibilities in a hierarchy.
The model extends the hierarchy levels defined in \emph{IEC 62264-1} by adding the concept \emph{Product} on the lowest level and the concept \emph{Connected World} at the top level, which goes beyond the boundaries of an individual factory.
Standards for smart factories are aligned with the IT layer of RAMI4.0 and located at different layers (Figure~\ref{fig:RAMIStandards}). Thus, the location of OPC UA is at the IT communication layer indicates that it standardizes data management, integration, and analytic processes.

\subsection{Standards Landscape for Smart Manufacturing Systems}
The \emph{National Institute of Standards and Technology} (NIST) has defined a landscape of standards with focus on Smart Manufacturing Systems~\cite{lu2016current}.
Two major classifications have been done in this work. 
First, the classification of standards regarding three manufacturing-related life-cycles: 1) product development life-cycle standards; 2) production system life-cycle standards; and 3) business cycle for supply chain management.
Second, the classification regarding the ISA 95 manufacturing pyramid, which classifies standards into five levels, i.e., from the device to the enterprise level.     
\subsection{National Smart Manufacturing Standards Architecture}
The Ministry of Industry and Information Technology of China, in a joint effort with the Standardization Administration, created a report for defining a National Smart Manufacturing Standards Architecture~\cite{LiJTCLZ16}. 
The architecture comprises three dimensions: 
(1) Smart functions, which include resource elements, system integration, interconnect, and information convergence to new business models; 
(2) The life-cycle, comprising design, production, logistics, marketing and sales to service; and 
(3) Hierarchy levels, consisting of equipment, control, workshop, and enterprise to inter-enterprise collaboration.
Although this standardization framework is three-dimensional as RAMI4.0, the classification of the features of the standards is conducted based on different criteria. 
For instance, the OPC UA standards is classified in the hierarchy level in the field level. 
This situation also generates semantic interoperability conflicts between organizations that utilize OPC UA following these two divergent classifications.    

\subsection{Industrial Internet Reference Architecture}
The \emph{Industrial Internet Reference Architecture} (IIRA) is a standard-based open architecture for IoT-based systems.
To support the smart industry vision, IIRA defines a generic description and representation with a high level of abstraction.
It provides a framework comprising methods to design industrial internet systems, without making specific recommendations for standards that comprise these systems~\cite{IIRA2017}.
IIRA includes the industrial internet viewpoints, i.e., business, usage, functional, and implementation. 
These viewpoints provide an analysis of individual sets of IoT-based systems. 
Further, the \emph{Industrial Internet Connectivity Framework} (IICF) extends IIRA to map existing standards with different functional levels. 
These levels range from the physical, link, network, transport, framework, and the top level of distributed data interoperability and management. 
For instance, the Message Queuing Telemetry Transport (\emph{MQTT}) standard\furl{https://mqtt.org/} is mapped to the level of transport.
As can be observed, existing standards are described in various ways in these frameworks; this lack of a unified view of the classification of the standards ends up in semantic interoperability conflicts. 
In this article, we devise data management techniques able to integrate standards into the I40KG. 
Standards are described in I40KG using the \emph{STO} ontology and represented in a structured way. 
Moreover, relationships between standards and among standards and standardization frameworks are explicitly stated, as well as their main properties and characteristics. 
Further, semantics encoded in \emph{STO} enables the execution of a reasoning process that uncovers implicit relationships. 
The explicit representation of these relationships allow for stating mappings  between divergent characterizations of standards w.r.t different standardization frameworks, reducing thus semantic interoperability conflicts.

\subsection{Semantic Interoperability Conflicts}
\label{sec-inter}
In this subsection, we describe the Semantic Interoperability Conflicts (SIC).
\begin{figure*}[t]
  \centering    
  \includegraphics[width=1\linewidth]{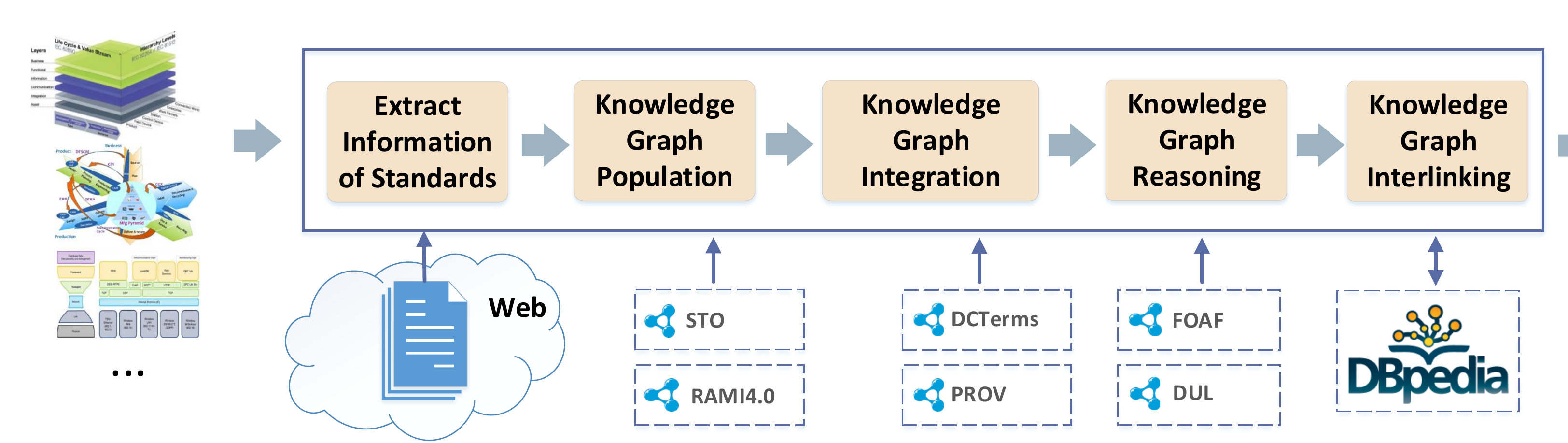}
  \caption{\textbf{Methodology for creating the I40KG}. I40 standards are received as input and the output is a graph representing relations between standards. \emph{STO} and existing vocabularies are utilized to describe known relations between standards. A reasoning process exploits the semantics encoded in \emph{STO} to infer new relations between standards. The linking with external KGs, e.g., \emph{DBpedia}, permits the enrichment of the I40KG. }
  \label{fig:stds-pipeline}
\end{figure*}
A SIC denotes differences in modeling and characterizing a real-world concept, e.g., a standard~\cite{JirkovskyOM17}. 
General semantic interoperability conflicts~\cite{bergman2006sources}, they can grouped into six main categories: 1) \textbf{Domain (SIC1)}: this interoperability conflict occurs when various interpretations of the same domain are represented. 2) \textbf{Schematic (SIC2)}: this interoperability conflict exists between data sources that are modeled using different schema. 3) \textbf{Granularity (SIC3)}: this interoperability conflict arises when various interpretations of the same domain are represented. 4) \textbf{Representation (SIC4)}: this interoperability conflict occurs when different representations are used to model the same concept.  5) \textbf{Missing Item (SIC5)}: this interoperability conflict appears whenever different items in distinct data sources are missing. 6) \textbf{Language (SIC6)}: this interoperability conflict appears whenever different languages are used to represent the data or metadata, i.e., schema.

\subsubsection{Semantic interoperability conflicts in Industry~4.0}
The semantic interoperability conflicts described in the previous section affect also the I40 domain, e.g., data integration.
In this work, we focus on semantic interoperability conflicts that occur between standardization frameworks and standards.
\textbf{Conflicts in standardization frameworks:}
We identify two main semantic interoperability conflicts impacting the ability to exchange information between standardization frameworks: 1) the categorization of same concepts and functions typically important to realize smart manufacturing is described in different layers across divergent standardization frameworks; 
2) the categorization of same standards is made in distinct dimensions and layers in different standardization frameworks (cf. Figure \ref{fig:reference-architectures}).
\textbf{Conflicts in standards:}
Common semantic interoperability conflicts between standards include: 1) \textit{Homonyms}: same terms are described with distinct meanings in different standards; e.g., the term \emph{Resource} is described in \emph{ISO 15704} as: ``An enterprise entity that provides some or all of the capabilities required by the execution of an enterprise and/or business process";
whereas in \emph{ISO 10303} as: ``something that may be described in terms of a behaviour, a capability, or a performance measure that is pertinent to a given process". 
2) \textit{Acronyms}: different abbreviations are used to refer to the same standard; e.g., IEC 62541 and OPC UA. 
3) \textit{Synonyms}: distinct names are utilized to express the same meaning, e.g., an \emph{InternalElement} in AML describes the same meaning as an \emph{Object} in OPC UA.

\section{Related Work}
\label{relatedWork}
In this section, we describe state-of-the-art approaches related to this work.
For better understanding, we classify them into three main categories: \begin{inparaenum}[1)]
\item generic semantic data integration approaches;
\item ontology-based approaches for representing I40 standards; and finally \item techniques and methods to solve semantic interoperability conflicts among standards and standardization frameworks.  
\end{inparaenum}
Generic semantic data integration approaches aim at solving semantic interoperability conflicts independently of the domain. 
Collarana \emph{et al.}~\cite{CollaranaGRV0A17} introduce MINTE, an integration framework that collects and integrates data from heterogeneous sources into a KG. 
MINTE implements semantic integration techniques that rely on the concept of RDF molecules to represent the meaning of data.
This approach also implements fusion policies for merging the RDF molecules and resolve semantic interoperability conflicts between the heterogeneous sources.
The aforementioned generic approaches for semantic data integration act upon structured data. 
On the contrary, the I40KG is built by manually extracting the knowledge of standards and standardization frameworks from non-structured data. Additionally, specific semantic interoperability conflicts for standards and standardization frameworks are identified and taken into account when building the I40KG.
Ontology-based approaches for representing Industry~4.0 standards are concerned with the use of the semantics of ontologies to express the shared knowledge of the standards. 
A framework to analyze the IoT standardization landscape is presented in~\cite{darmoisiot}.
In this work, smart manufacturing is considered as a vertical dimension of IoT.
A standard database classifying standards is defined in an abstract way, e.g., generic and domain-specific standards.
Finally, general gaps of standards and their functions related to IoT are described. 
Steinmetz \emph{et al.}~\cite{steinmetz2018using} outline an ontology based approach for integrating IoT based information in I40.
The presented work discuss how industrial assets can be represented by means of an ontology.
The aim is to perform semantic integration of the data that assets generate.
The performed integration enables the communication of the assets through an IoT middleware system. 
Existing ontology-based approaches for representing I40 standards suffer from several limitations.
First, no dedicated ontology is considered for semantically representing standards and standardization frameworks concepts and their associated metadata. 
Second, relations between standards are identified but not modeled by means of an ontology. 
The development of the I40KG is based on the semantic encoded in the \emph{STO} ontology. 
The \emph{STO} ontology covers the concepts of standards and standardization frameworks as well as the metadata associated with them, which is necessary for representing the knowledge in this domain.  
Existing works for solving semantic interoperability conflicts refer to the identification of standards and their alignment to a level or layer of certain standardization frameworks.
Lin \emph{et al.}~\cite{ramiirareport2017} present similarities and differences between the RAMI4.0 model and the IIRA architecture.
Based on the study of these similarities and differences authors proposed a functional alignment between layers in RAMI4.0 with the functional domains and crosscutting functions in IIRA.  
Additionally, in this work, the IICF framework, which extends IIRA, outlines layers of IoT and identify standards for each one of these layers. 
Furthermore, the layers in RAMI4.0 are aligned to the IICF layers. 
For example, while RAMI4.0 specifies OPC UA as the core standard for connecting manufacturing products, equipment and process software, IICF also specifies OPC UA and adds other three standards, i.e., TCP/UDP/IP, TSN.
Lu \emph{et al.}~\cite{lu2016current} describe a standardization landscape for smart manufacturing systems. 
The landscape is built upon relations of standards with products, production systems, and business life-cycle dimensions.
The landscape is also described in terms of standards organizations as well as types of standards acting in each of the three dimensions.
\begin{figure*}[t!]
  \centering    
  \includegraphics[width=1\linewidth,height=0.3\linewidth]{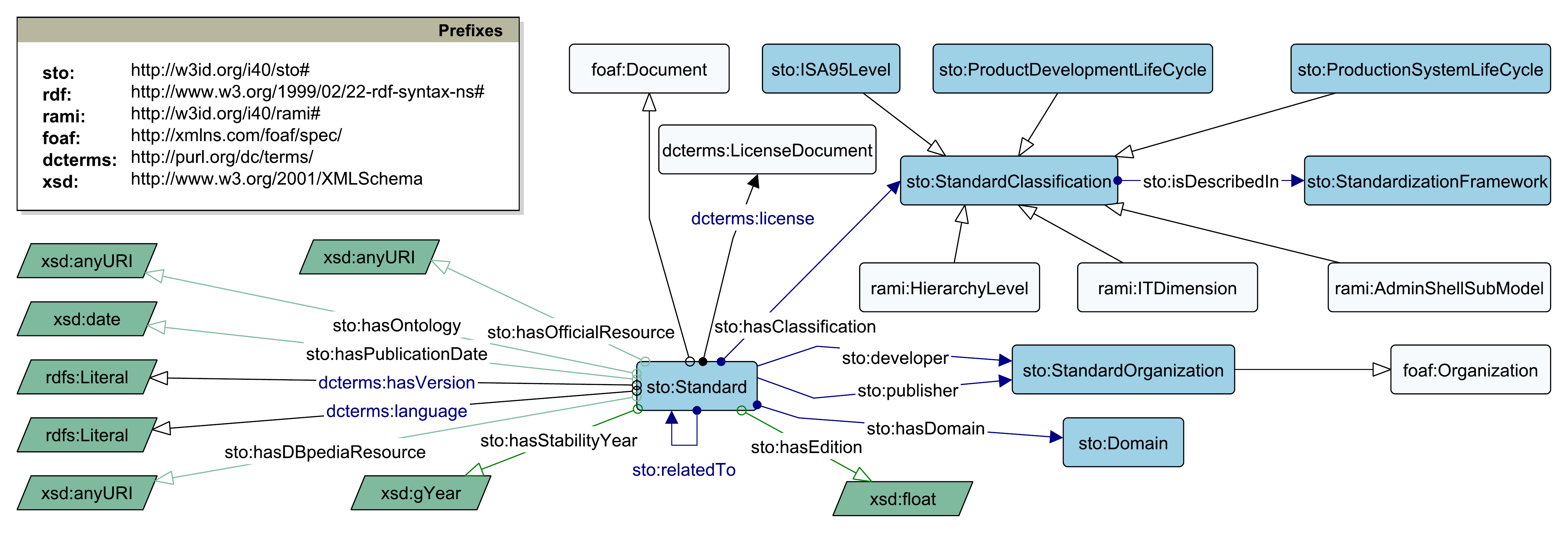}
  \caption{\textbf{Core classes and properties of the Standards Ontology (\emph{STO})}. Classes of \emph{STO} are depicted in blue. White classes represent reused classes from FOAF, DCTERMS, and RAMI4.0 ontologies. Reused properties are drawn in blue, e.g., \texttt{dcterms:language}. Green rectangles depicts datatype properties. Classes and properties are used to semantically describe standards and standardization frameworks in I40 scenarios.} 
  \label{fig:STO}
\end{figure*}
Recent works for semantically integrating I40 related standards are presented by us in~\cite{alligatorekaw2016} and ~\cite{Grangel-Gonzalez18}.
The first approach, Alligator, combines the power of Datalog and ontologies to semantically integrate KGs while solving semantic interoperability conflicts between I40 related standards. 
The second approach, SemCPS, uses probabilistic techniques to resolve existing semantic interoperability conflicts and integrate KG of I40 related standards.
Many shortcomings can be outlined by investigating aforementioned methods. 
First, their focus is on identifying and classifying existing relations and semantic interoperability conflicts between standardization frameworks, e.g., ~\cite{ramiirareport2017,lu2016current,LiJTCLZ16}. 
In contrast, this article provides a methodological foundation for identifying existing semantic interoperability conflicts between standards and standardization frameworks and include them as a part of the I40KG. 
Second, the rest of the approaches only target the integration between the information models of the standards, e.g., \cite{HodgesGR17,alligatorekaw2016,Grangel-Gonzalez18}.
Conversely, the I40KG targets to resolve semantic interoperability conflicts not only between standards but also between standards and standardization frameworks.
The aforementioned approaches comprise several limitations to resolve semantic interoperability conflicts among standards as well as among standards and standardization frameworks. 
Contrary, in this article, we provide a methodological foundation for the creation and refinement of the I40KG.
Further, a characterization of existing semantic interoperability conflicts that are common in this scenarios. 
Finally, the I40KG comprises the semantic metadata of the main concepts in the domain allowing for the identification and solution of semantic interoperability conflicts.

\section{I40 Knowledge Graph creation}
\label{sec-methodology}
In this section, we present our methodology for the creation of the I40KG. 
Figure~\ref{fig:stds-pipeline} shows the five steps comprising this methodology: i) Extract Information of Standards; ii) Knowledge Graph Population; iii) Knowledge Graph Integration; iv) Knowledge Graph Reasoning; and v) Knowledge Graph Interlinking. 
Initially, we describe the Standards Ontology, as one of the key components used along all of the steps of our methodology.
Next, each step of the methodology is elaborated in detail.
\subsection{The Standards Ontology}
\label{ontology}
The Standards Ontology (\emph{STO}) semantically describe I40 standards as well as their relations.
In addition, the STO ontology encodes four main standardization frameworks for I40 that classify I40 standards.  
\emph{STO} is modeled according to the best practices for building ontologies.
In this regard, classes and properties from well-known ontologies are reused, e.g., the \emph{PROV} ontology for describing provenance of entities; the \emph{FOAF} ontology for representing and linking documents and agents, e.g., persons, organizations; the \emph{DCTERMS} ontology for documenting metadata, such as licenses, as well as the \emph{RAMI4.0} ontology for linking standards with \emph{RAMI4.0} concepts.

\subsection{Information Extraction}
To extract information of standards, we investigated documents related to the standardization frameworks for I40.
We started by combining terms such as ``Industry 4.0'', ``Reference Architectures'', ``Standardization Landscape'', and ``Standards''.
Following the selected criteria, a list of relevant documents for standardization frameworks used to describe standards is compiled, e.g., the RAMI4.0 Model~\cite{ramiModel} and the IIRA architecture~\cite{IIRA2017}.
For each found framework, a corresponding RDF molecule with the name and URI of the framework is created and added to the I40KG.
Next, we continue with retrieving responsible organizations that publish or develop standards by using the following terms ``Standardization Organizations'', ``Industry 4.0'', ``Standards'', and ``Reference Architectures''.
These terms are introduced on various search engines and the first top ten documents are chosen for further investigation. 
As a result, a list of 30 standardization organizations is compiled. 
For each organization in the list, an RDF molecule with the name of the standardization organization and its URI is introduced in the I40KG.
Furthermore, to obtain more information about organizations, we used their name and search for each one of them.
In case for a specific organization an official page exists, its link is added to the respective RDF molecule.
From its official page, the acronym and the formation date are retrieved.  
Additionally, if the links to Wikipedia and DBpedia exist, they are also added to the molecule. 
In the next step, a list of standards that are described in the retrieved documents of the standardization frameworks is computed.
Standards are searched on the retrieved documents using the pattern ``publishing organization'' and ``numeric value'', e.g., IEC as the organization and 62541 as the numeric value.
The obtained mappings enable the link of existing standards to respective standardization frameworks.
We further investigated standardization frameworks comprising information about classifications of standards.
RAMI4.0 classifies standards according to three general dimensions, i.e., \emph{IT}, Life-Cycle and Value Stream, and the Hierarchy Level.
For example, the \emph{IT} dimension of RAMI4.0 represents a number of different layers, starting from the \emph{Asset} to the \emph{Business} layer (cf. Figure~\ref{fig:RAMIStandards}). 
Furthermore, various standards are mapped to the specific layers of this dimension.
The Administration Shell concept is part of the RAMI4.0 framework. 
Due to its importance for the I40 domain, we considered it as another standardization framework used to classify standards. 
In this case, it delineates how standards are linked to submodels.
The submodels enclose the different functions required by a particular asset, e.g., Identification, Communication, or Engineering~\cite{ramiShellStructure}.
The Identification submodel is aligned with the \emph{ISO 29005} standard, whereas the Communication submodel with the \emph{IEC 61784 Fieldbus Profiles}.
For instance, the Engineering submodel is aligned with standards such as \emph{IEC 61360}, \emph{IEC 61987}, and \emph{{eCl@ss}}.
The step of extracting information of standards from unstructured data sources allows for representing this knowledge using the \emph{STO} ontology and encoding it in I40KG. 
Hence, the conflicts \textbf{SIC1}, \textbf{SIC2}, and \textbf{SIC4}, which exist across standardization frameworks and standards, are resolved during this step.
\subsection{Knowledge Graph Population}
The population of I40KG with standards and standardization frameworks is realized according to the terminology defined in the \emph{STO} ontology.
To perform this step, we consider the concepts of RDF molecules and RDF molecule templates (RDF-MTs).
An RDF-MT is an abstract representation of the set of properties associated with an RDF class, and all links between the class with other RDF classes~\cite{EndrisGLMVA17}. 
Instances of an RDF-MT correspond to RDF molecules in a KG.
The RDF-MTs describe the relations between classes in a KG and the classes from the external KGs to which it is linked.
A unique URI is defined for each name of standard to prevent the duplication of RDF molecules.
For each standard, in case many results are obtained, the maximum number of 20 documents is selected. 
By following this method, a total number of 220 documents of different types are retrieved, i.e., technical reports (12), white papers (6), scientific articles (28), standard specifications (165), technical presentations (7), and technical papers (2).
The period in which these documents were published ranges from December 2002 to July 2017. 
For each one of the documents, the properties of the standard RDF-MT are examined. 
The obtained values are used to build one RDF molecule for each standard. 
Furthermore, the mappings to the standardization frameworks are added to each RDF molecule for a respective standard. 
The linking to external KGs is a common method for KG completion~\cite{Paulheim17}.
Therefore, to utilize the external knowledge, we performed a linking procedure of I40KG to \emph{DBpedia}~\cite{AuerBKLCI07}, by inspective the names of the standards. 
In case that the name exists in Wikipedia, e.g., https://en.wikipedia.org/wiki/IEC\_61131, it is also present in DBpedia with the same name, e.g., http://dbpedia.org/page/IEC\_61131.
Then, the property \texttt{sto:hasDBpediaResource} is used to connect a particular standard to its equivalent representation in DBpedia. 
The output is a list of the RDF molecules including the mappings of standards and standardization frameworks. 
The property \texttt{sto:relatedTo} describes that two standards are mentioned in one document but no explicit relation is defined. In case that explicit relations are defined, their name is encoded as in the form of respective properties.
\begin{lstlisting}[style=turtle, caption={\textbf{RDF-based description of the molecule of the OPC UA standard}}, label={lst:opcUAresource}]
@prefix sto:  <https://w3id.org/i40/sto#> .
@prefix rami: <https://w3id.org/i40/rami#> .
@prefix rdfs: <http://www.w3.org/2000/01/rdf-schema#> .
sto:OPC_UA                a sto:Standard;
 rdfs:label               "OPC UA"@en;
 rdfs:comment             "International standard for ..."@en;
 sto:hasTag               "OPC UA"@en;
 sto:hasPublisher         sto:OPC_Foundation;
 sto:hasDeveloper         sto:OPC_Foundation;
 sto:hasDBpediaResource   <http://dbpedia.org/page/
                          OPC_Unified_Architecture>;
 sto:hasOfficialResource  <https://opcfoundation.org/
                          about/opc-technologies/opc-ua/>;
 sto:hasWikipediaArticle  <https://en.wikipedia.org/wiki/
                          OPC_Unified_Architecture>;
 sto:isInteroperableWith  sto:AML;
 sto:integratesWith       sto:IEC_61499;
 sto:hasDomain            sto:M2MCommunication;
 sto:hasClassification    rami:Communication; sto:FrameworkLevel ;
 dcterms:license          sto:GPLv2.                   
\end{lstlisting}
The OPC UA standard is described with respect to relations with other standards as well as by classifying it regarding to distinct standardization frameworks~\ref{lst:opcUAresource}.
Currently, the I40KG comprises more than 200 standards and more than 25 standard organizations.
Moreover, 66 direct relations between standards are encoded as a part of the KG.
I40KG is publicly available and can be expanded by the community with interest in I40 as well as domain experts in this topic by directly accessing it on Github\footnote{https://github.com/i40-Tools/StandardsOntology}.
\subsection{Knowledge Graph Integration}
The knowledge integration step enables semantic definition of the connections between instances in I40KG to resolving interoperability conflicts.  
For example, there are cases when similar standards share the same meaning but are named differently, which causes a semantic interoperability conflict \textbf{SIC1} between those standards.
This applies to the \emph{IEC 62541} standard, that is actually the OPC UA standard published by the IEC organization but known with a different name in the IEC publication. 
In this case, additional properties for the RDF standard molecule of IEC 62541 are considered, e.g., available languages, the technical committee, and the stability date.
These properties are extracted from the official website of the IEC\furl{https://webstore.iec.ch/publication/21996}.
These standards, i.e., OPC UA and IEC 62541, are named differently depending on the organizations, but they refer to the same standard.
Thus, they are considered as equivalent entities and integrated into the I40KG.
\subsection{Knowledge Graph Reasoning}
One of the main motivations to create I40KG is to encode the knowledge of I40 and to study the existing relations among them.
The reasoning step is performed with the aim to unveil new knowledge from the implicitly stated relations. 
This step enables retrieving explicitly-defined relations in I40KG as well as those that are inferred.
For instance, relations of AML (IEC 62714) and OPC UA (IEC 62541), OPC UA (IEC 62541) and IEC 61499 are explicitly defined in the KG. 
Based on this fact, the relation between AML and IEC 61499 is inferred.
The existence of this relation is checked and validated in the literature and is of importance for the domain~\cite{pang2010iec}. 
Furthermore, a number of different relations between standards in the I40KG can be retrieved. 
The result of this query reports on 103 relations among the standards, and 266 new relations inferred when the symmetric and transitive properties are considered by the inference process.
Several graph metrics are computed over the two graphs, i.e., original and inferred graph, respectively, to analyze the connectivity and relationships discovered during the reasoning step. 
The number of edges increases from 66 to 227, indicating the discovery of the new relations among standards. 
The graph density--the fraction of the number of potential connections in the graph that are actual connections--  increases slightly; it goes from 0.025 to 0.068.
This implies a slight improvement of connectivity among the standards. 
Values of transitivity are augmented, i.e., from 0.12 to 0.732.
This indicates an increment of the possibility that a relation between two standards in the graph is transitive.
The clustering coefficient also increases from 0.085 to 0.389.
These results highlight an increase in the degree to which the standards with the same connections, tend to cluster together.
The inferred graph also becomes more centralized, i.e., from 0.128 to 0.237. 
These findings reveal the importance of standards within the I40KG. 
For instance, IEC 62541 (OPC UA) with a value of centrality of 0.8, seems to be more important than ISO 20922 with a value of 0.58.

\subsection{Knowledge Graph Interlinking}
The interlinking step enables interconnecting I40KG with KGs in the Linked Open Data Cloud (LOD)~\cite{berners2009linked}, i.e., DBpedia.
To accomplish this step, every standard in I40KG is surveyed.
A link is automatically created whenever the same instance of a particular standard exists in both KGs, in I40KG and DBpedia. 
Next, by using this link, knowledge from DBpedia is extracted and added to I40KG.
During this step it is possible to discover new knowledge of standards and standard organizations. 
Additionally, the knowledge can be validated based on what is encoded in I40KG.
To this end, queries are executed on DBpedia. 
These queries are based on the linked instances of standards and organizations.
Finally, the I40KG is enriched with additional facts, i.e., new classes, properties and instances, which are added after the KG interlinking process. 
Any DBpedia property that does not add semantic value to I40KG, e.g., \texttt{dbo:wikiPageID}, \texttt{dbo:wikiPageRevisionID}\footnote{dbo is the namespace of http://dbpedia.org/ontology/} is omitted.
\section{Evaluation}
\label{sec:evaluation}
We empirically study the quality and accuracy of the proposed KG driven approach.
Precision is used as a quality metric to evaluate our approach.
We are particularly interested in answering the following research questions:
\begin{inparaenum}[\bf {\bf RQ}1\upshape)]
\item Can a \emph{knowledge graph driven} approach allow for the discovery of valid relations among standards?;
\item Is the proposed \emph{knowledge graph interlinking step} able to discover new knowledge in terms of classes, properties, and instances?
\end{inparaenum}
In the following sections, we detail the experiments set up, results and discuss the lessons learned. 
\subsection{\textbf{RQ1} - Discovering relations between standards}
To answer \textbf{RQ1}, we validate the existence and correctness of the discovered relations between standards in I40KG, against a new set of I40 documents retrieved from the Web.
Two criteria are considered to determine whether the discovered relations can be evaluated as true: 1) direct relations between standards; and 2) in case standards appeared in the same document indicating the similar goal, e.g., I40, Smart Manufacturing. 
A new set of I40 documents on the Web is generated by exploring a number of various search engines, such as Google, Google Scholar, Bing, Yahoo, and Ask.
To perform the search of the 266 discovered relations, a combination of the names and tags of the standards is used, e.g., ``AML" and ``IEC 61499".
As a result, a total number of 148 documents are retrieved comprising research papers (66), white papers (8), standard specifications (17), technical reports (48), technical presentations (6), and thesis (3).
These documents are different from those employed to create I40KG.
A total number of 188 relations out of 266 new relations are validated as correct, resulting on a Precision value equal to 0.71.
Accordingly, we are able to positively answer \textbf{RQ1}.
A total of 266 new relations out of 66 initial are inferred and 188 are validated by searching into different types of documents; the method reaches a 0.71 value of Precision. 
\subsection{\textbf{RQ2} -  Discovering knowledge through knowledge graph interlinking}
\begin{figure*}[htb!]
\begin{subfigure}{0.31\textwidth}
\includegraphics[height=.73\linewidth]{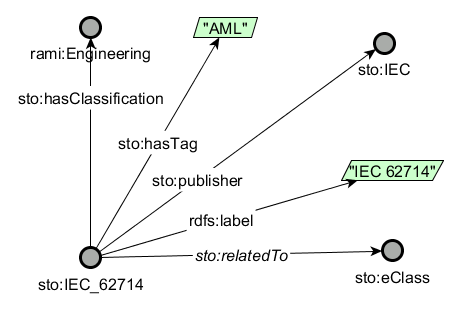}
\caption{Initial IEC 62714 molecule}
\label{fig:molecule_template_a}
\end{subfigure}
\begin{subfigure}{0.32\textwidth}
\includegraphics[height=.72\linewidth]{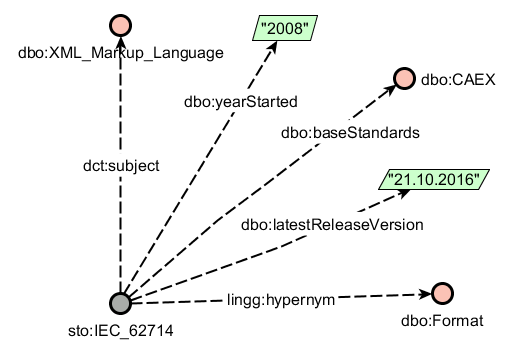}
\caption{New connections}
\label{fig:molecule_template_b}
\end{subfigure}
\begin{subfigure}{0.33\textwidth}
\includegraphics[height=.7\linewidth]{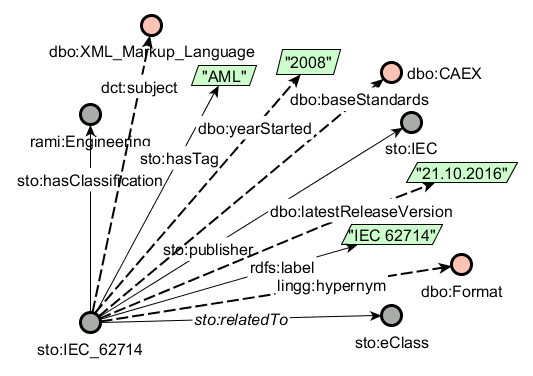}
\caption{Final molecule enrichment}
\label{fig:molecule_template_c}
\end{subfigure}
\caption{\textbf{Example of the RDF molecule for the IEC 62714 (AML) standard before and after the KG interlinking step}. Figure \ref{fig:molecule_template_a} shows the basic RDF molecule of one standard in the I40KG, i.e., \texttt{sto:IEC\_62714} before the KG interlinking step. Figure \ref{fig:molecule_template_b} depicts five of the new connections, properties, instances, and literal values that are incorporated for the \texttt{sto:IEC\_62714} standard. Figure \ref{fig:molecule_template_c} illustrates the complete molecule for \texttt{sto:IEC\_62714} after the KG interlinking step.} 
\label{fig:aml_mol_template}
\end{figure*}

To measure the effectiveness of the KG approach after the linking step, the two following criteria are studied: the number of new class linkings, and the number of new properties.
These criteria are computed by considering instances of standards and standard organizations. 
\begin{table}[!htbp]
\scriptsize
\caption{\textbf{Precision values for properties related to the RDF-MTs of standards and standard organizations after the KG interlinking step}. 
Precision values for new class linkings, i.e., \texttt{rdf:type} and the total number of new properties after executing the KG interlinking step are reported. Instances of standards (\textbf{Std}) and standard organizations (\textbf{Org}) are examined.} 
\label{tab-precision}
\begin{tabular}{|l|l|l|l|l|}
\hline
\textbf{Criteria} & \textbf{Total Std} & \textbf{Total Org} & \textbf{Precision Std} & \textbf{Precision Org}  \\ \hline
New class linkings & 93 & 108 & 0.66 & 0.90 \\ \hline
New properties & 35 & 39 & 0.97 & 0.96 \\\hline
\end{tabular}
\vspace{-4mm}
\end{table}
\begin{inparaenum}
\item \emph{Number of new class linkings}. This refers to the number of new classes that are automatically added to the instance as types.
For example, the DBpedia class \emph{Industry XML Specific Standards} is added to the AML standard as one of its types. 
A possible duplicate consideration of classes is avoided, i.e., a given class is considered only once for a specific standard. 
\item \emph{Number of new properties}. This refers to the number of new properties that are automatically added to each instance of a standard.
For instance, the \emph{IEC 62714} standard is enriched with the following new properties:
\texttt{dcterms:subject}, and 
\texttt{dbo:yearStarted}, which are not considered in I40KG (cf. Figure \ref{fig:aml_mol_template}).
In this case, two properties are assessed for the number of new properties. 
A manual inspection is performed over the obtained properties. 
Any particular property which is not defined in DBpedia as an \texttt{rdf:property} is counted as a false positive. 
Furthermore, whenever a property does not add value nor have a description associated, is also not considered. 
\end{inparaenum}
SPARQL queries, particularly designed for each criterion, are executed on top of the enriched I40KG to materialize the results of both new classes and properties. 
Table~\ref{tab-metrics} reports on the applied queries to the number of standards (73) and standard organization instances (22) which contain a link to DBpedia.
We observe from the table that new knowledge for standards and organizations is discovered. 
Precision increases by up to 0.96, suggesting thus that the accuracy of the discovered relations is increased by up to 96\%. 
\subsection{Effectiveness of the graph interlinking step}
To measure the quality of the generated connections in the I40KG, we evaluate the KG interlinking step. 
This evaluation is important since the automatic linking of RDF KGs can be prone to different types of errors, i.e., syntactic, logical, and semantic errors.
For example, the RDF triple \texttt{sto:SCOR rdf:type dbo:Person} in DBpedia indicates that the standard SCOR is a person. Links from I4.0KG to the resource \texttt{sto:SCOR} in DBpedia also introduce this error in I4.0KG and affect the quality of the represented relations. 
During the evaluation of the effectiveness of the interlinking  process, the validity of the relations is measured in terms of Precision.    

\begin{table}[!htbp]
\scriptsize
\caption{\textbf{Precision values for properties of the RDF-MT of standard and standard organizations after the KG interlinking step}. Values of three properties are observed, i.e., \texttt{owl:sameAs} links, \texttt{dcterms:subject}, and \texttt{lingg:hypernym}.
\textbf{Precision} is computed based on the values of these properties for instances of standards (\textbf{Std}) and standardization organizations (\textbf{Org}).} 
\label{tab-metrics} 
\begin{tabular}{|l|l|l|p{1cm}|p{1cm}|}
\hline
\textbf{Criteria} & \textbf{Total Std} & \textbf{Total Org} & \textbf{Precision Std} & \textbf{Precision Org}  \\ \hline
\texttt{owl:sameAs}  & 301  & 217 & 0.98 & 0.91 \\ \hline
\texttt{dcterms:subject} & 144  & 115 & 0.98 & 0.96 \\ \hline
\texttt{lingg:hypernym} & 44  & 36 & 0.61 & 0.85 \\ \hline
\end{tabular}
\end{table}

The properties are related to the values of the instances to which they are linked. 
Second, using the \texttt{rdf:type} property, standards or organizations are enriched with new class linkings. 
However, there are cases when new statements are wrongly generated.
For example, the following statement: \texttt{sto:SCOR rdf:type dbo:Person} is considered as a semantic error and the value is marked as false, since the \emph{Supply Chain Operation Reference Model} \texttt{sto:SCOR} is a standard and not a person. 
Similarly, in case that SCOR is classified as a model, i.e., \texttt{sto:SCOR rdf:type yago:Model}\footnote{yago, prefix of the YAGO KG: http://yago-knowledge.org/resource/} is considered as a true statement. 
Consequently, the existing classification in DBpedia is correct and marked as a true positive for the interlinking step. 
Values of the \texttt{dcterms:subject}, are also inspected regarding semantic errors according to its definition which describes the topic of a given resource\footnote{http://dublincore.org/usage/terms/history/\#subjectT-001}.
Based on this definition, values are observed by checking whether they correctly represent a topic or not.
A similar process is conducted for the property \texttt{lingg:hypernym}, where we checked whether the linked resource is a hypernym, i.e., a broader classification of the instance. 
For instance, the standard \emph{B2MML}, which is defined as an XML implementation of the ANSI/ISA-95 family of standards, has \emph{implementation} as its hypernym.
In this case, the new link is correct, and we counted it as a true positive; otherwise as a false positive. 
With respect to \texttt{owl:sameAs}, links to other instances representing the same semantics are revised.
The values of the four above properties are inspected for each enriched instance during the knowledge graph interlinking step, i.e., 103 instances of standards, and 23 of standard organizations. 
Table \ref{tab-precision} reports on the computed values of Precision. 

\subsection{RDF Molecule Templates}

RDF-MTs describe the relations between classes in a knowledge graph and the classes of the KGs to which it is linked.
Instances of an RDF-MT correspond to RDF molecules in a KG.
Table~\ref{tab-metrics} shows the computed Precision for the values of the properties added to the RDF-MT of standard (35) and standard organization (86) after the KG interlinking step, respectively.
Precision values for studied properties, i.e., \texttt{owl:sameAs}, \texttt{dcterms:subject}, and \texttt{lingg:hypernym} of the RDF-MT of standard and standard organizations after the KG interlinking step. 
Precision is computed based on the values of these properties for instances of standards and standardization organizations.
Relations for the RDF-MTs based on these properties can be validated by up to 98\%. 
Figure~\ref{fig:aml_mol_template} depicts an example of the standard RDF-MT based on the RDF molecule of the AML (IEC 62714) standard. 
The left side of the figure shows the basic standard RDF-MT, exemplified with the AML standard before the KG interlinking step. 
Figure \ref{fig:molecule_template_b} depicts eight of the new connections, i.e., properties, instances, and literal values that are incorporated to the standard RDF-MT, whereas Figure~\ref{fig:molecule_template_c} illustrates the standard RDF-MT after the KG interlinking step. 
Only five properties are included for space reasons but the standard RDF-MT is extended with 35 additional properties to represent each standard. 
Likewise, the RDF-MT for standard organizations is enriched with 88 new properties. 
The graph comprises 221 RDF-MTs and 259 intra- and inter-knowledge graph links. 
YAGO and DBpedia are the most utilized KGs. 
As expected, two RDF-MTs, i.e., \texttt{sto:Standard} and \texttt{sto:StandardOrganization} are the source of most of the generated links.
The analysis is performed before and after the KG interlink step to study the I40KG w.r.t the connectivity and discovered relationships. 
The total number of 221 RDF-MTs with 259 links are generated based on the initial 43.
Particularly, we observe that the graph density decreases, i.e., from 0.025 to 0.009, which can due to the many types of new RDF-MTs added after the KG interlinked step.
The clustering coefficient increases, i.e., from 0.072 to 0.108, indicating that the degree to which the RDF-MTs in I40KG graph tend to cluster together is increased to 33\%.
Values of transitivity are rather low and experiment a decrease, i.e., from 0.017 to 0.0057.
This result can be explained by considering that while more RDF-MTs are added, they are not connected in a transitive manner. Further, values of centralization are increased to 15\%.
As expected, the most important RDF-MTs, which are the central concepts of the I40KG are the \texttt{sto:Standard} (0.45) and \texttt{sto:StandardizationFramework} (0.41).

\section{Conclusion and Future Work}
\label{conclFutureWork}

In this article, a Knowledge Graph of Industry~4.0 related standards (I40KG) and the Standard Ontology (\emph{STO}) for the semantic description of standards and their relations are developed.
Further, a methodology for building knowledge graphs of Industry~4.0 related standards is presented. 
We investigated existing reference models like RAMI4.0 and NIST, and populated I40KG with descriptions of more than 200 standards, more than 25 standardization organizations, and 100 relations between the standards.
Finally, the I40KG has been linked to existing knowledge graphs such as DBpedia and automated reasoning has been implemented to reveal implicit relations between standards as well as mappings across standardization frameworks.
We analyze both the number of discovered relations among standards and the accuracy of these relations. 
Observed results indicate that both, reasoning and linking processes enable for increasing the connectivity in the knowledge graph by up to 80\%, whilst up to 96\% of the relations can be validated.
These outcomes suggest that integrating standards and standardization frameworks into the I40KG enable the resolution of semantic interoperability conflicts, empowering thus the \emph{communication} in smart factories. 
We hope that this work contributes to a crucial step in realizing the Industry~4.0 vision, since it requires not only standards governing individual aspects but needs to consider semantics in the relations among standards as well as standards and standardization frameworks.
As for the future work, we envision to expand the knowledge graph of Industry~4.0 standards as well as the semantic annotations of the standards by means of the ontology.
Additionally, we aim to consider categorizations of standards available in other reference architectures.
Further, we plan to exploit the semantics encoded in the \emph{STO} ontology as well as in the I40KG to determine semantics similarities between standards based on their terms. 

\section*{Acknowledgements}
This work has been partially supported by the EU H2020 funded projects PLATOON (GA No. 872592) and by the project BRIDGE-US (GA No. 01DD19005) supported by the Federal Ministry of Education and Research, Germany.

\printbibliography

@inproceedings{Grangel-Gonzalez18,
  author    = {Irl{\'{a}}n Grangel{-}Gonz{\'{a}}lez and Lavdim Halilaj and Maria{-}Esther Vidal and Omar Rana and Steffen Lohmann and S{\"{o}}ren Auer and Andreas W. M{\"{u}}ller},
  title     = {Knowledge Graphs for Semantically Integrating Cyber-Physical Systems},
  booktitle = {Database and Expert Systems Applications - 29th International Conference, {DEXA}, Regensburg, Germany, September 3-6, Proceedings,
               Part {I}},
  pages     = {184--199},
  year      = {2018}
}

@article{steinmetz2018using,
  title={Using Ontology and Standard Middleware for integrating {I}o{T} based in the {I}ndustry 4.0},
  author={Steinmetz, Charles and Rettberg, Achim and Ribeiro, Fab{\'\i}ola Gon{\c{c}}alves C and Schroeder, Greyce and Soares, Michel S and Pereira, Carlos E},
  journal={IFAC-PapersOnLine},
  volume={51},
  number={10},
  pages={169--174},
  year={2018},
  publisher={Elsevier}
}

@inproceedings{CollaranaGRV0A17,
  author    = {Diego Collarana and
               Mikhail Galkin and
               Ignacio Traverso Rib{\'{o}}n and Maria{-}Esther Vidal and
               Christoph Lange and
               S{\"{o}}ren Auer},
  title     = {{MINTE:} semantically integrating {RDF} graphs},
  booktitle = {7th Int. Conf. on Web Intelligence, Mining and Semantics, {WIMS}, Amantea, Italy, June 19-22},
  pages     = {22:1--22:11},
  year      = {2017}
}

@inproceedings{EndrisGLMVA17,
  author    = {Kemele M. Endris and
               Mikhail Galkin and
               Ioanna Lytra and
               Mohamed Nadjib Mami and
               Maria{-}Esther Vidal and
               S{\"{o}}ren Auer},
  title     = {{MULDER:} Querying the Linked Data Web by Bridging {RDF} Molecule Templates},
  booktitle = {Database and Expert Systems Applications - 28th Int. Conf., {DEXA} Lyon, France, August 28-31, Proceedings, Part {I}},
  pages     = {3--18},
  year      = {2017}
}

@article{HodgesGR17,
	author    = {Jack Hodges and Kimberly Garc{\'{i}}a and Steven Ray},
	title     = {Semantic {D}evelopment and {I}ntegration of {S}tandards for {A}doption and {I}nteroperability},
	journal   = {{IEEE} Computer},
	volume    = {50},
	number    = {11},
	pages     = {26--36},
	year      = {2017}
}

@TechReport{ramiirareport2017,
  author        = {Shi-Wan Lin and Brett Murphy and Erich Clauer and Ulrich Loewen and Ralf Neubert and Gerd Bachmann and Madhusudan Pai and Martin Hankel},
  title         = {Reference {A}rchitectural {M}odel {I}ndustrie 4.0 ({RAMI 4.0})},
  institution   = {Industrial {I}nternet {C}onsortium and {P}lattform {I}ndustrie 4.0},
  year          = {2017},
  date          = {2017-12-05}
}

@article{Paulheim17,
  author    = {Heiko Paulheim},
  title     = {Knowledge graph refinement: {A} survey of approaches and evaluation methods},
  journal   = {Semantic Web},
  volume    = {8},
  number    = {3},
  pages     = {489--508},
  year      = {2017},
  url       = {https://doi.org/10.3233/SW-160218}
}

@techreport{IIRA2017,
  title={The {I}ndustrial {I}nternet of {T}hings {V}olume {G1}: {R}eference {A}rchitecture},
  author={Shi-Wan Lin and Bradford Miller and Jacques Durand and Graham Bleakley and Amine Chigani and Robert Martin and Brett Murphy and Mark Crawford},
  type={White Paper},
  institution={Industrial {I}nternet {C}onsortium},
  year={2017}
}

@article{JirkovskyOM17,
  author    = {V{\'{a}}clav Jirkovsk{\'{y}} and
               Marek Obitko and
               Vladim{\'{i}}r Mar{\'{i}}k},
  title     = {Understanding Data Heterogeneity in the Context of Cyber-Physical
               Systems Integration},
  journal   = {{IEEE} Trans. Industrial Informatics},
  volume    = {13},
  number    = {2},
  pages     = {660--667},
  year      = {2017}
}

@inproceedings{LuMF15,
  author    = {Yan Lu and
               K. C. Morris and
               Simon Frechette},
  title     = {Standards landscape and directions for smart manufacturing systems},
  booktitle = {{IEEE} {I}nternational {C}onference on {A}utomation {S}cience and {E}ngineering,
               {CASE}, Gothenburg, Sweden, August 24-28},
  pages     = {998--1005},
  year      = {2015}
}

@inproceedings{LiJTCLZ16,
  author    = {Qing Li and
               Hongzhen Jiang and
               Qianlin Tang and
               Yaotang Chen and
               Jun Li and
               Jian Zhou},
  title     = {Smart Manufacturing Standardization: Reference Model and Standards
               Framework},
  booktitle = {On the Move to Meaningful Internet Systems: {OTM} Workshops},
  pages     = {16--25},
  year      = {2016},
  publisher = {Springer},  
}

@article{weyer2015towards,
  title={Towards {I}ndustry 4.0-{S}tandardization as the crucial challenge for highly modular, multi-vendor production systems},
  author={Weyer, Stephan and Schmitt, Mathias and Ohmer, Moritz and Gorecky, Dominic},
  journal={IFAC-PapersOnLine},
  volume={48},
  number={3},
  pages={579--584},
  year={2015},
  publisher={Elsevier}
}

@article{darmoisiot,
  title={Io{T} Standards--State-of-the-Art Analysis},
  author={Darmois, Emmanuel and Elloumi, Omar and Guillemin, Patrick and Moretto, Philippe},
  journal={Digitising the Industry Internet of Things Connecting the Physical, Digital and Virtual Worlds},
  pages={978--87},
  year = {2012}
}

@article{lu2016current,
  title={Current standards landscape for smart manufacturing systems},
  author={Lu, Yan and Morris, KC and Frechette, Simon},
  journal={National Institute of Standards and Technology},
  volume={8107},
  year={2016}
}

@inproceedings{alligatorekaw2016,
  author    = {Irl\'an Grangel-Gonz\'alez and
               Diego Collarana and
               Lavdim Halilaj and
               Steffen Lohmann and
               Christoph Lange and
               Maria-Esther Vidal and
               S{\"o}ren Auer
               },
  title     = {Alligator: A Deductive Approach for the Integration of {I}ndustry 4.0 Standards},
  booktitle = {20th Int. Conf. on Knowledge Engineering and Knowledge Management, {EKAW}},
  year      = {2016},
  pages     = {272--287},  
}

@inproceedings{Grangel-Gonzalez17,
  author    = {Irl{\'{a}}n Grangel{-}Gonz{\'{a}}lez and
               Paul Baptista and
               Lavdim Halilaj and
               Steffen Lohmann and
               Maria{-}Esther Vidal and
               Christian Mader and
               S{\"{o}}ren Auer},
  title     = {The industry 4.0 standards landscape from a semantic integration perspective},
  booktitle = {22nd {IEEE} Int. Conf. on Emerging Technologies and Factory Automation, {ETFA}, Limassol, Cyprus, September 12-15},
  pages     = {1--8},
  year      = {2017}
}

@techreport{ramiModel,
  title={Reference {A}rchitecture {M}odel {I}ndustrie 4.0 ({RAMI4.0})},
  year={2015},
	author={Peter Adolphs and Heinz Bedenbender and Dagmar Dirzus and Martin Ehlich and Ulrich Epple and Martin Hankel and Roland Heidel and Michael Hoffmeister and Haimo Huhle and Bernd K{\"{a}}rcher and Heiko Koziolek and Reinhold Pichler and Stefan Pollmeier and Frank Schewe and Armin Walter and Bernd Waser and Martin Wollschlaeger},
	institution={ZVEI and VDI},
    type = {Status Report}
}

@techreport{ramiShellStructure,
  title={Structure of the Administration Shell},
  year={2016},
    author={Peter Adolphs and S\"oren Auer and Meik Billmann and Martin Hankel and Roland Heidel and Michael Hoffmeister and Haimo Huhle and Michael Jochem and Markus Kiele and Gunther Koschnick and Heiko Koziolek and Lukas Linke and Reinhold Pichler and Frank Schewe and Karsten Schneider and Bernd Waser},
	institution={ZVEI and VDI},
    type = {Status Report}
}

@article{bergman2006sources,
  title={Sources and classification of semantic heterogeneities},
  author={Bergman, M},
  journal={Web {B}log: {AI3}-{A}daptive {I}nformation, {A}daptive {I}nnovation, {A}daptive {I}nfrastructure},
  year={2006}
}

@inproceedings{pang2010iec,
  title={{IEC} 61499 function block implementation of {I}ntelligent {M}echatronic {C}omponent},
  author={Pang, Cheng and Vyatkin, Valeriy},
  booktitle={8th IEEE {I}nt. {C}onf. on Industrial {I}nformatics},
  pages={1124--1129},
  year={2010},
  organization={IEEE}
}

@article{berners2009linked,
  title={Linked data--the story so far},
  author={Tim Berners-Lee and Christian Bizer and Tom Heath},
  journal={International {J}ournal on {S}emantic {W}eb and {I}nformation {S}ystems},
  volume={5},
  number={3},
  pages={1--22},
  year={2009}
}

@inproceedings{AuerBKLCI07,
  author    = {S{\"{o}}ren Auer and
               Christian Bizer and
               Georgi Kobilarov and
               Jens Lehmann and
               Richard Cyganiak and
               Zachary G. Ives},
  title     = {DBpedia: {A} Nucleus for a Web of Open Data},
  booktitle = {The Semantic Web, 6th Int. Semantic Web Conference, 2nd Asian Semantic Web Conference, {ISWC} + {ASWC}, Busan, Korea, November 11-15.},
  pages     = {722--735},
  year      = {2007}
}
\end{document}